\documentclass[
twocolumn,
aps,
prb,
amsmath,amssymb,
showpacs,
superscriptaddress
]{revtex4-2}

\usepackage[utf8]{inputenc}
\usepackage{epsfig}
\usepackage{float}
\usepackage{color}
\usepackage{bm}
\usepackage{amssymb,amsmath}
\usepackage{graphicx}
\usepackage[colorlinks=true,urlcolor=blue,citecolor=blue]{hyperref}

\usepackage{color}

\begin{document}

\title{Magnetic field-tuned magnetic order and metamagnetic criticality in non-stoichiometric CeAuBi$_2$}

\author{Halyna Hodovanets} 
\email{halyna.hodovanets@mst.edu}
\author{Hyunsoo Kim} 
\affiliation{Maryland Quantum Materials Center, Department of Physics, University of Maryland, College Park, MD 20742, USA}
\affiliation{Department of Physics, Missouri University of Science and Technology, Rolla, Missouri 65409, USA}

\author{Tristin Metz} 
\author{Yasuyuki Nakajima} 
\author{Christopher J. Eckberg} 
\author{Kefeng Wang} 
\author{Jie Yong} 
\author{Shanta R. Saha}
\affiliation{Maryland Quantum Materials Center, Department of Physics, University of Maryland, College Park, MD 20742, USA}
\author{David Graf}
\affiliation{National High Magnetic Field Laboratory, Florida State University, Tallahassee, Florida 32313, USA}

\author{Nicholas P. Butch} 
\affiliation{Maryland Quantum Materials Center, Department of Physics, University of Maryland, College Park, MD 20742, USA}
\affiliation{NIST Center for Neutron Research, National Institute of Standards and Technology, Gaithersburg, Maryland 20899, USA}

\author{Thomas Vojta} 
\affiliation{Department of Physics, Missouri University of Science and Technology, Rolla, Missouri 65409, USA}

\author{Johnpierre Paglione}
\email{paglione@umd.edu}
\affiliation{Maryland Quantum Materials Center, Department of Physics, University of Maryland, College Park, MD 20742, USA}
\affiliation{Canadian Institute for Advanced Research, Toronto, Ontario M5G 1Z8, Canada}

\begin{abstract}
We present a detailed study of magnetization, resistivity, heat capacity, and X-ray and neutron powder diffraction measurements performed on single crystals of non-stoichiometric CeAuBi$_2$, Au deficiency 18$\%$, a strongly correlated antiferromagnet with N\'eel temperature $T_N$=13.2 K. 
Field-dependent magnetization measurements reveal a large magnetic anisotropy at low temperatures with an easy axis along the crystallographic $c$-axis, in which direction a spin-flop transition exhibits strong features in magnetization, specific heat, and resistivity at $H_c = 75$ kOe. The constructed temperature-field phase diagram connects this transition to the suppression of magnetic order, which evolves from a  second-order nature into a first-order transition that bifurcates at the spin-flop transition into three transitions below 1 K. 
The smoothed nature of the metamagnetic transitions in non-stoichiometric CeAuBi$_2$ is well described by an Ising model with weak quenched disorder, suggesting that the presence of Au vacancies is sufficient to smear the complex metamagnetic behavior and tune the critical behavior of magnetic order.

\end{abstract}

\maketitle

\section{Introduction}

The 4$f$ intermetallic compounds display a rich variety of possible ground states due to the competition between the Ruderman-Kittel-Kasuya-Yosida (RKKY) and Kondo interactions.\cite{Stewart1984,Stewart2001,Stewart2006} Both mechanisms depend on the exchange interaction $J$ between the local $f$-moment and the conduction electrons as well as the density of states at the Fermi surface, $N(E_F)$.\cite{Doniach1977} With applied pressure, chemical substitution, or magnetic field, further evolution of these ground states can be achieved. For example, the magnetic second-order ferromagnetic (FM) or antiferromagnetic (AFM) transition evolves through a continuous phase transition into a quantum critical point (QCP) at $T$ = 0 with new ground states in the vicinity of this QCP \cite{Doniach1977,Hewson1997,Coleman2007} If the QCP is avoided because the magnetic transition becomes of the first order, a quantum phase transition (QPT) occurs upon suppression of magnetism with magnetic field resulting in very rich and complex $T-H$ phase diagrams. 

In the Ce$TX_2$ (where $T$ = Cu, Ag, and Au, and $X$ = Sb and Bi) family, all compounds order antiferromagnetically with sizable magnetic anisotropy and the moments along an easy $c$-axis with the exceptions of CeCuSb$_2$, that has a very small magnetic anisotropy\cite{Thamizhavel2003a} and no $T-H$ phase diagram reported, and CeAgSb$_2$, that orders ferromagnetically and for which $c$-axis is a hard axis\cite{Myers1999a}. CeCuBi$_2$,\cite{Thamizhavel2003} CeAuSb$_2$,\cite{Lorenzer2013} and CeAgBi$_2$\cite{Thamizhavel2003,Thomas2016} all display very rich and complex $T-H$ phase diagrams for \textbf{H}$\|$\textbf{c} with few metamagnetic transitions. Five metamagnetic transitions observed in the $T-H$ phase diagram of CeAgBi$_2$ (\textbf{H}$\|$\textbf{c}) were explained as a consequence of the competing exchange interaction and anisotropy.\cite{Thomas2016} The anisotropy and crystalline electric field (CEF) parameters vary strongly with \textit{T} within the family being the largest for CeAuBi$_2$.\cite{Thamizhavel2003,Sidorov2003,Adriano2014,Myers1999a,Araki2003,Seibel2015,Adriano2015,Thomas2016}  

A similar collection of metamagnetic transitions was recently observed in nearly stoichiometric CeAuBi$_2$\cite{Piva2020}, which exhibits an antiferromagnetic order below $T_N$=19~K that increases with applied pressure up to 21~K at 23~kbar \cite{Piva2020}, suggesting a rich field-temperature phase space that is sensitive to external tuning parameters. 
Here, we study the effect of stoichiometry on the AFM ordering temperature and its evolution with the magnetic field, finding that an 18$\%$ deficiency of Au reduces the N\'eel temperature to $T_N\sim$ 13 K and tunes the complex metamagnetic behavior to very low temperatures.
We provide a comprehensive study of the temperature-field evolution using magnetization, heat capacity, resistivity, and powder diffraction techniques. We construct a detailed phase diagram mapping the relation between AFM order and the field-induced metamagnetic transitions, revealing a spin-flop transition for fields applied along the magnetic easy-axis, and construct an Ising model to show that the sharp metamagnetic transitions are smeared out by weak quenched disorder due to Au vacancies. A change in the nature of the in-field magnetic transition from continuous to first-order at the lowest temperatures is understood in the framework of the Ising model.

\section{Experimental details}

Single crystals of nominally stoichiometric CeAuBi$_2$ were grown using the high temperature flux method.\cite{Canfield1992,Canfield2010,Canfield2001} Chunks of Ce (99.8$\%$ purity, AlfaAesar), powder of Au (99.96$\%$ purity, AlfaAesar) and chunks of Bi (99.999$\%$ purity, AlfaAesar) in the ratio of 5:5:90 were placed in the alumina crucible, sealed in the quartz ampule under partial Ar pressure, heated to 1100~$^o$C, held at that temperature for 2 h, cooled down at $\sim$ 3 deg/min to 670~$^o$C at which temperature the excess Bi was decanted with the help of the centrifuge. The quartz ampules were opened in the nitrogen-filled glovebox. The samples before and after measurements were stored in the glovebox with nitrogen atmosphere as well. Single crystals ground in air immediately oxidized. The outer layer of the single crystals degrades relatively quickly if left in the air. However, if the sample is cleaved, the cleaved surface shows no degradation. Some surface degradation can also be seen in the samples stored in the glovebox for a long time. Magnetization measurements, made twice on the same sample stored in the glovebox, the second time four months later, did not show any significant change in T$_N$, critical field of a metamagnetic transition, and saturated moment observed at 140 kOe. Samples showing visible degradation also show a complete resistive superconducting transition with $T_c$ and $H_{c2}$ similar to what was observed in CeNi$_{1-x}$Bi$_2$\cite{Lin2013}. Energy dispersive X-ray spectroscopy (EDS) analysis indicated that CeAuBi$_2$ single crystals are stoichiometric; however, wavelength dispersive X-ray spectroscopy (WDS) showed that the crystals were slightly Au deficient. From the EDS, the ratios of Ce:Au:Bi were 1:1.04:2.19 with an error bar of 2 standard deviation of 0.21 for Au and 0.40 for Bi. The larger ratio of Bi is most likely due to the excess Bi on the surface of the crystals. The WDS ratios were Ce:Au:Bi = 1:0.82:1.99 with error bars of 0.09 and 0.28 for Au and Bi, respectively. We will refer to our samples in this manuscript as CeAuBi$_2$ bearing in mind that they are 18$\%$ Au deficient.
All measurements were made, unless otherwise specified, on the samples from the same batch. Attempts to synthesize single crystals of LaAuBi$_2$ were not successful. 

The Bruker D8 Advance powder diffractometer (Cu radiation) and the Rigaku MiniFlex diffractometer (Cu radiation) were used to collect X-ray diffraction patterns. The single crystals were ground in the glovebox (nitrogen atmosphere) and placed either in the airtight sample holder (Bruker D8) or covered with a Kapton tape (Rigaku MiniFlex) to avoid powder oxidation.

For the neutron diffraction experiment, six batches of single crystals were grown together with the same ratio of elements (except that chunks of Au (99.999$\%$ purity) were used instead of Au powder) and temperature profile. A crystal from each batch was measured using the Quantum Design VSM option with \textbf{H}$\|$\textbf{c}. $T_N$ for these samples was estimated to be 12.86 $\pm$ 0.05 K and is lower than the $T_N$ of the samples used for other measurements. For the neutron diffraction experiment we assumed that the slight variation of the lattice parameters and Au concentration should not affect the origin or type of magnetic ordering. The single crystals were ground and sealed in a 50 mm vanadium container with a diameter of 6.0 mm inside a dry He-filled glovebox. A closed-cycle He refrigerator was used for temperature control. Data were collected at 20 K (above the AFM ordering) and at 4, 6, 8, 10, 12, and 13 K (in the AFM state). Neutron powder diffraction data were collected using the BT-1 32 detector neutron powder diffractometer at the NCNR, NBSR. A Cu(311) monochromator with $\lambda$ = 1.5403(2) {\AA} was used. Data were collected in the range of $3-155\,^{\circ}$ 2-Theta with a step size of $0.05\,^{\circ}$. The instrument is described in the NCNR WWW site (http://www.ncnr.nist.gov/). Rietveld refinement of structural and magnetic phases was performed using GSAS\cite{Toby2013} and version 2K of the program SARA{\it h}-Representational Analysis\cite{Wills2000}.

Quantum Design Physical Property Measurement System PPMS DynaCool$^{TM}$ was used to perform temperature- and field-dependent resistivity, magnetization, and specific heat measurements. Resistivity measurements were made in a standard four-probe geometry, $\it ac$ technique ($\it f$ = 16 Hz, $\it I$ = 1 mA). Electrical contact to the samples was made with silver wires attached to the samples using silver paste which was allowed to cure at room temperature. The contact resistance at room temperature was less than 1 $\Omega$. The current was flowing along the naturally formed edge of the plates, current along a (100) direction of the \textit{ab}-pane. The distance between the midpoint of the two voltage contacts and the cross section area of the samples were used to calculate the resistivity of the samples. The magnetization was measured using the Vibrating Sample Magnetometer Option (VSM). The samples were mounted with the help of GE varnish. The contribution of GE varnish to the $M(T)/H$ and $M(H)$ data was assumed to be negligible.

Temperature and field-dependent resistivity and specific heat measurements were extended to $\sim$ 50 mk and up to 150 kOe in a dilution refrigerator. Field-dependent resistivity measurements of up to 315 kOe were performed at the National High Magnetic Field Laboratory in Tallahasse. 

For the theoretical model, following Ref.\cite{Thomas2016}, we considered the Hamiltonian on a magnetic unit cell that contains 16 Ce atoms arranged in two bilayers. Each bilayer can be understood as a buckled square lattice with nearest neighbor and next-nearest neighbor interactions $J$ and $J_\perp$, respectively. The bilayers are coupled to each other via $J'$. To find the low-temperature magnetic state for a given set of parameters ($J$, $J'$, $J_\perp$, $\Delta$, and $h$), we numerically minimized the total energy of the magnetic unit cell w.r.t. the classical spin variables  $\mathbf{S}_i$. The minimization started from completely random $\mathbf{S}_i$ and employed the NEWUOA software for unconstrained optimization \cite{Powell2006}. As optimization sometimes gets stuck in metastable states, we repeated the optimization a large number of times for each parameter set (typically 100 to 300) and identified the lowest energy among the resulting states.

To study the effects of quenched disorder, we generated a large number ($10^4$ to $10^5$) of independent samples, i.e., realizations of the magnetic unit cell with different realizations of the random interactions. We then applied the optimization procedure described above to each of the samples and averaged the resulting magnetization values. Note that this is an approximate description of the macroscopic disordered system, as it effectively implies that the same random unit cell is repeated in all directions. To improve this approximation one could consider larger magnetic unit cells, at the expense of a much larger numerical effort. We anticipate that this would lead to minor quantitative changes of the results, but not to qualitative ones.

%----------------------------------------------------------------------------------------------------------
%----------------------------------------------------------------------------------------------------------
\section{Results}
\subsection{X-ray diffraction}
\begin{figure}[tbh]
\centering
\includegraphics[width=1\linewidth]{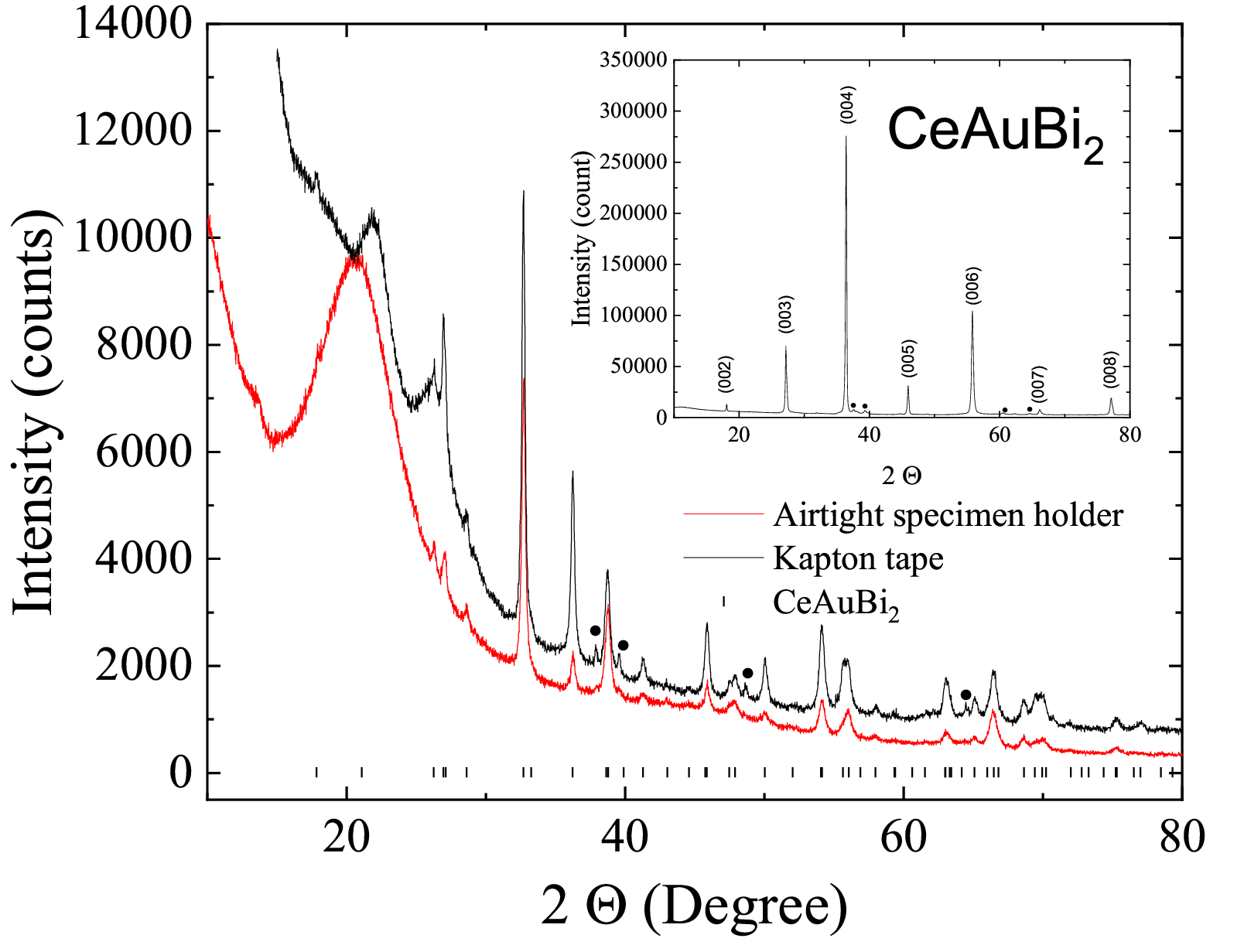}
\caption{\footnotesize Powder X-ray diffraction patterns of ground CeAuBi$_2$ single crystals. Broad peaks around 2$\Theta$ = 20 degrees are due to Kapton tape and acrylic glass (airtight container). The inset: X-ray diffraction pattern of the single crystal of CeAuBi$_2$. The few low-intensity peaks marked with dots can be associated with Bi flux.}
\label{XRD}

\end{figure}
\begin{figure*}[t]
\centering
\includegraphics[width=1\linewidth]{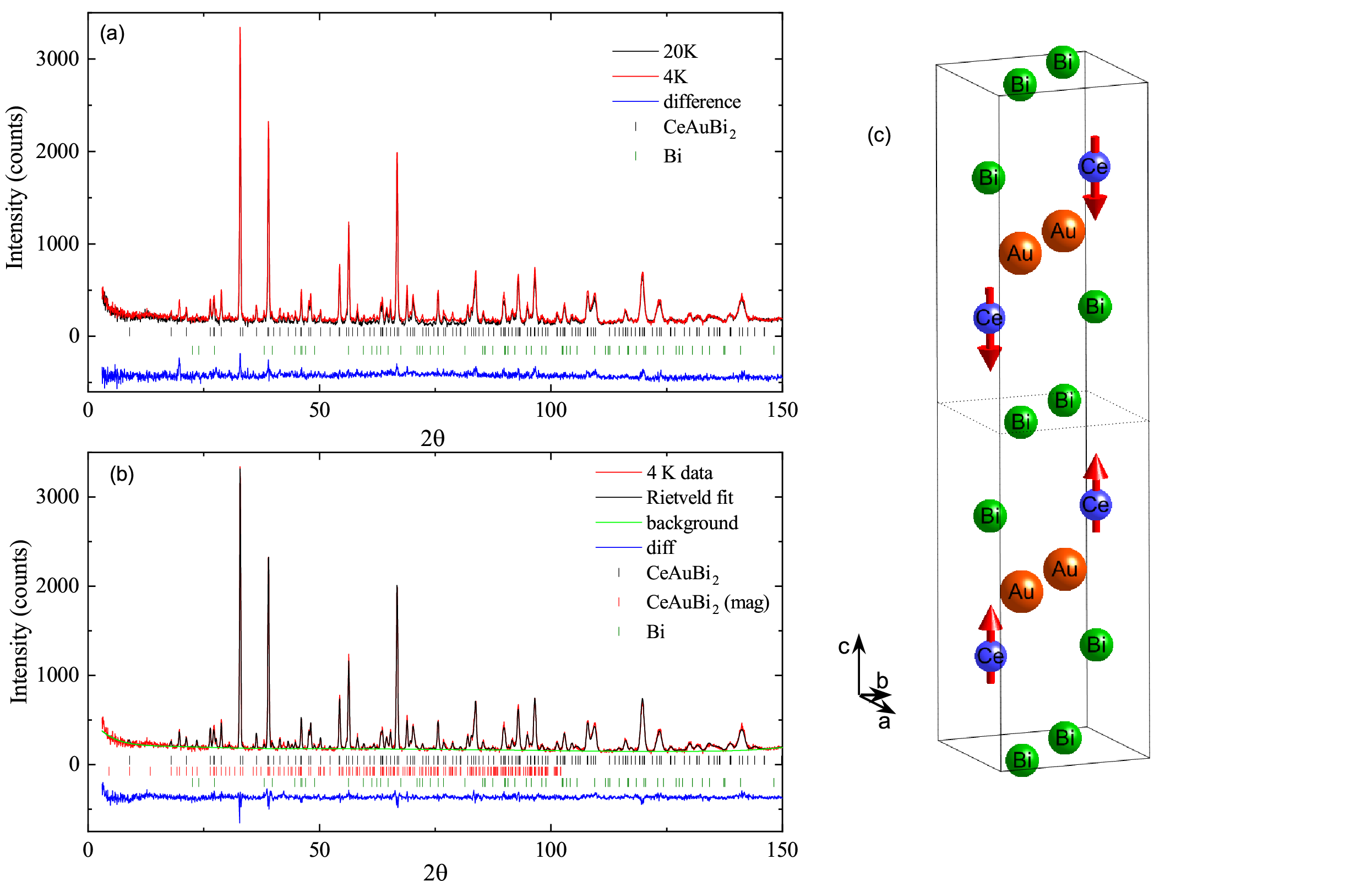}
\caption{\footnotesize (a) Powder neutron diffraction patterns of ground CeAuBi$_2$ single crystals at 20 and 4 K. The pattern in blue shows the difference between the two patterns, i.e. the magnetic contribution. (b) Powder neutron diffraction pattern at 4 K with Rietveld fit (black curve) and the difference between them (blue curve). (c) Schematic representation of the magnetic structure (solid lines) of CeAuBi$_2$. The magnetic unit cell is twice the crystallographic unit cell (represented by a dashed line) along the $c$-direction. (d) Magnetic moment refined from the neutron diffraction data as a function of temperature.}
\label{NU}
\end{figure*}

 CeAuBi$_2$ crystallizes in the HfCuSi$_2$ type structure (space group P4/nmm).\cite{Seibel2015} X-ray diffraction patterns collected on the ground single crystals of CeAuBi$_2$ are shown in Fig. \ref{XRD}. Both, the airtight sample holder and Kapton tape, contribute to the broad amorphous peak at $\sim$ 2$\Theta$ = 20 degree and a larger than usual background to the diffraction patterns. We also collected the X-ray pattern on a single crystal of CeAuBi$_2$, the inset of a Fig. \ref{XRD}. Since the $c$-axis is perpendicular to the naturally formed plates, only reflections that belong to (00\textit{l}) appear in the X-ray pattern. The lattice parameters obtained form the LeBail fit are: \textit{a} = 4.63 $\pm$ 0.01 {\AA} and \textit{c} = 9.89	$\pm$ 0.02 {\AA} (D8, polycrystals), \textit{a} = 4.64 $\pm$ 0.01 {\AA} and \textit{c} = 9.89	$\pm$ 0.02 {\AA} (MiniFlex, polycrystals), and \textit{c} = 9.89	$\pm$ 0.02 {\AA} (Rigaku, single crystal). The \textit{c} lattice parameter is slightly smaller than that in the recently published work.\cite{Adriano2015,Seibel2015}

\subsection{Neutron diffraction}

\begin{table}[tbh]
\caption{Basis vectors (BV) for the space group P 4/nmm:2 with 
${\bf k}=( 0,~ 0,~ 0.5)$.The 
decomposition of the magnetic representation for 
the $Ce$ site 
$( 0.25,~ 0.25,~ 0.26327)$ is 
$\Gamma_{Mag}=1\Gamma_{2}^{1}+1\Gamma_{3}^{1}+1\Gamma_{9}^{2}+1\Gamma_{10}^{2}$. The atoms of the nonprimitive 
basis are defined according to 
1: $( 0.25,~ 0.25,~ 0.26327)$, 2: $( 0.75,~ 0.75,~ 0.73673)$.}
\begin{ruledtabular}
\scalebox{0.8}{\begin{tabular}{ccc||cccccc}
  IR  &  BV  &  Atom & \multicolumn{6}{c}{BV components}\\
      &      &             &$m_{\|a}$ & $m_{\|b}$ & $m_{\|c}$ &$im_{\|a}$ & $im_{\|b}$ & $im_{\|c}$ \\
\hline
$\Gamma_{2}$ & $\bf\psi_{1}$ &      1 &      0 &      0 &      8 &      0 &      0 &      0  \\
             &              &      2 &      0 &      0 &      8 &      0 &      0 &      0  \\
$\Gamma_{3}$ & $\bf\psi_{2}$ &      1 &      0 &      0 &      8 &      0 &      0 &      0  \\
             &              &      2 &      0 &      0 &     -8 &      0 &      0 &      0  \\
$\Gamma_{9}$ & $\bf\psi_{3}$ &      1 &      4 &      0 &      0 &      0 &      0 &      0  \\
             &              &      2 &     -4 &      0 &      0 &      0 &      0 &      0  \\
             & $\bf\psi_{4}$ &      1 &      0 &     -4 &      0 &      0 &      0 &      0  \\
             &              &      2 &      0 &      4 &      0 &      0 &      0 &      0  \\
$\Gamma_{10}$ & $\bf\psi_{5}$ &      1 &      4 &      0 &      0 &      0 &      0 &      0  \\
             &              &      2 &      4 &      0 &      0 &      0 &      0 &      0  \\
             & $\bf\psi_{6}$ &      1 &      0 &     -4 &      0 &      0 &      0 &      0  \\
             &              &      2 &      0 &     -4 &      0 &      0 &      0 &      0  \\
\end{tabular}}
\end{ruledtabular}
\label{bv}
\end{table}

\begin{figure*}[tbh]
\centering
\includegraphics[width=0.75\linewidth]{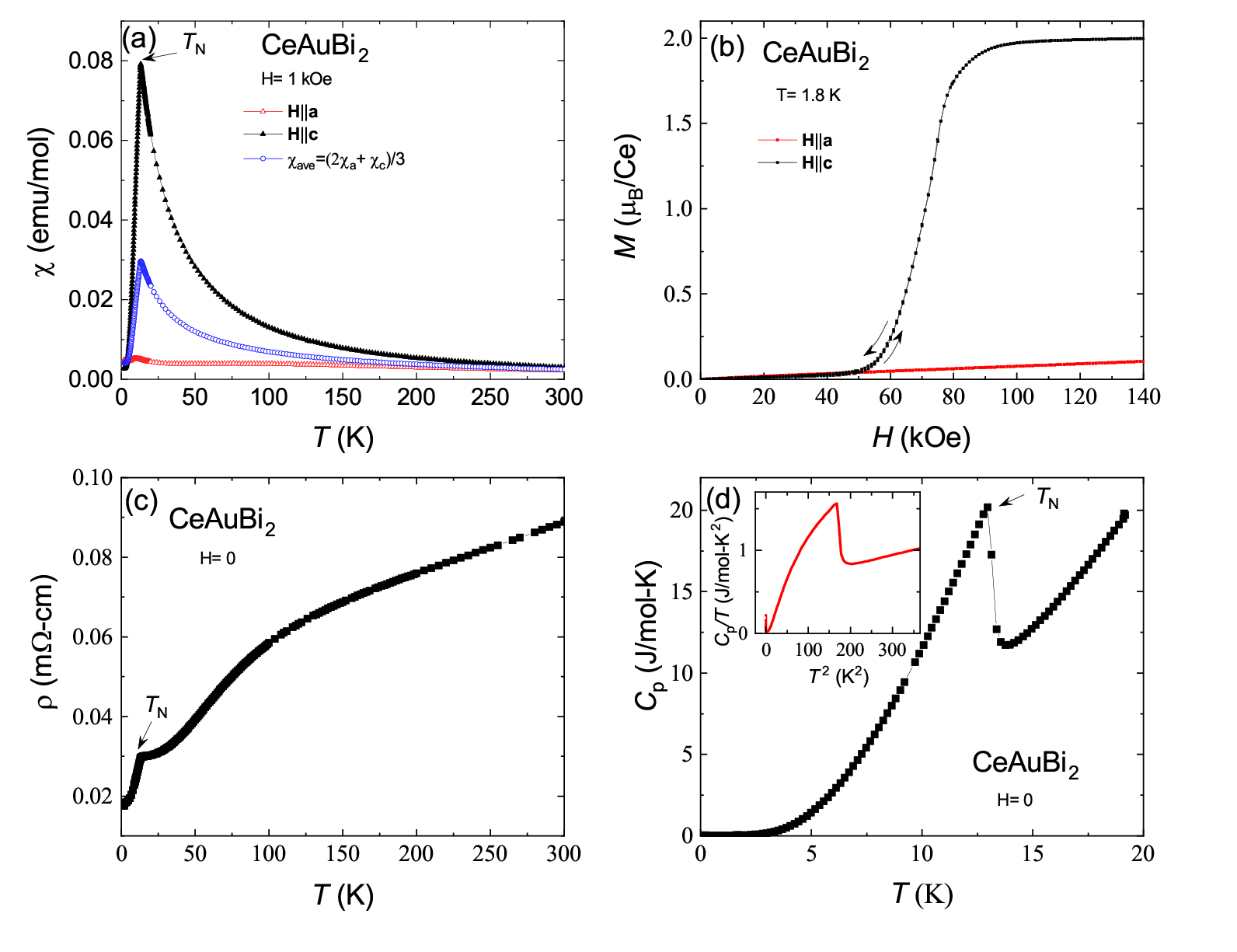}
\caption{\footnotesize (a) Temperature-dependent susceptibility of CeAuBi$_2$ single crystal for \textbf{H}$\|\textbf{a}$, \textbf{H}$\|\textbf{c}$ and the polycrystalline average $\chi_{ave}$=(2$\chi_a$+$\chi_c$)/3. %The inset shows d($\chi T$)/d$T$ as a function of temperature with 
The arrow denotes the AFM transition. (b) Field-dependent magnetization of CeAuBi$_2$ single crystal at 1.8 K for \textbf{H}$\|\textbf{a}$ and \textbf{H}$\|\textbf{c}$. (c) Temperature-dependent resistivity of CeAuBi$_2$ single crystal. (d) Temperature-dependent specific heat of CeAuBi$_2$. The inset shows $C_p/T$ vs $T^2$ data.}
\label{p1}
\end{figure*}

The neutron diffraction patterns at 20 and 4 K corresponding to $T>T_N$ and $T<T_N$, respectively, are shown in Fig. \ref{NU} (a). The difference between these two patterns, shown in blue, indicates the magnetic contribution due to the magnetic order at 4 K. At 20 K, all peaks can be indexed to CeAuBi$_2$ with $\sim$3 wt$\%$ of the secondary phase of Bi. The Au conteny was refined to be 0.88$\pm$0.01. The lattice parameters obtained from the Rietveld refinement are $a$ = 4.6147 {\AA} and $c$ = 9.876 {\AA}. This information was used to fit the nuclear contribution to the 4 K neutron data. Magnetic refinement at 4 K was performed based on the P1 space group with the propagation vector \textbf{k} = (0 0 $\frac{1}{2}$), Fig. \ref{NU}(c).  The crystal structure of CeAuBi$_2$ before the phase transition is described in the space group P4/nmm:2 (\#129:2). This space group involves 1 centering operations and 16 symmetry operations. Of these symmetry operations, 16 leave the propagation {\bf k} invariant or transform it into an equivalent vector, and form the little group $G_{\mathbf k}$. The magnetic representation of a crystallographic site can then be decomposed in terms of the irreducible representations (IRs) of $G_{\mathbf k}$:

\begin{equation}
 \Gamma_{Mag}=\sum_{\nu} n_{\nu}\Gamma_{\nu}^{\mu}
 \label{magnetic_representation}
\end{equation}

\noindent where n$_\nu$ is the number of times that the IR $\Gamma_\nu$ of order $\mu$ appears in the magnetic representation $\Gamma_{Mag}$ for the chosen crystallographic site.
The decomposition of the magnetic representation for the $Ce$ site $( .25,~ .25,~ .26327)$ is 
$\Gamma_{Mag}=0\Gamma_{1}^{1}+1\Gamma_{2}^{1}+1\Gamma_{3}^{1}+0\Gamma_{4}^{1}+0\Gamma_{5}^{1}+0\Gamma_{6}^{1}+0\Gamma_{7}^{1}+0\Gamma_{8}^{1}+1\Gamma_{9}^{2}+1\Gamma_{10}^{2}=1\Gamma_{2}^{1}+1\Gamma_{3}^{1}+1\Gamma_{9}^{2}+1\Gamma_{10}^{2}$. The decomposition of the magnetic representation $\Gamma_{Mag}$ in terms of the non-zero IRs of $G_{\bf k}$ for Ce crystallographic site, and their associated basis vectors, $\bf{\psi_n}$, are given in Table \ref{bv}. $\Gamma_{9}$ and $\Gamma_{10}$ IRs represent the magnetic moment in the \textit{ab}-plane which, as will be shown below, contradicts the magnetization measurements, where the \textit{c}-axis is the easy axis. Therefore, the refinement of the magnetic structure was not done based on those IRs. The basis vector for $\Gamma_{2}$ representation that gives (+-- --+) moments orientation along the \textit{c}-axis does not have a valid Shubnikov point group. This leaves $\Gamma_{3}$ representation that has the magnetic moment along the \textit{c}-axis with (++-- --) moment orientations and this IR was used for the magnetic structure refinement. The schematic chemical and magnetic unit cells are shown in Fig. \ref{NU}(c). The magnetic unit cell is equal to the chemical unit cell doubled along the \textit{c}-axis. The magnetic moment direction within the magnetic unit cell is shown with the red arrows. The magnetic structure is the same as that reported for CeCuBi$_2$\cite{Adriano2014} and CeAuBi$_2$\cite{Piva2020}. The refined magnetic moment was found to be $\mu$ = 1.92$\pm$0.03 $\mu_B$ and is consistent with the saturated moment at 4 K observed in \textit{M(H)} data shown below. We did not observe any change in the magnetic structure in the AFM state. The refined magnetic moment as a function of temperature is shown in Fig. \ref{NU}(d).

%----------------------------------------------------------------------------------------------------------
\subsection{Basic physical properties}

The temperature-dependent magnetic susceptibility data measured at 1 kOe for \textbf{H}$\|\textbf{a}$, \textbf{H}$\|\textbf{c}$, and the polycrystalline average taken as $\chi_{ave}$=(2$\chi_a$+$\chi_c$)/3 is shown in Fig. \ref{p1}(a). There is a strong anisotropy in the temperature-dependent magnetic susceptibility, $\chi_c>\chi_a$ for $T<$ 200 K. An arrow marks the N\'eel temperature $T_N$ = 13.2 $\pm$ 0.1 K, taken as a peak in $\chi_{ave}$. The ratio of $\chi_c/\chi_a\sim$ 16 at $T_N$. The Curie-Weiss fit of the polycrystalline average of susceptibility in the range 100 K $< T<$ 305 K resulted in $\mu_{eff}=2.52\mu_B$, which is in agreement with cerium in a trivalent state and $\Theta_p$=-13.6 K. However, if the modified Curie-Weiss fit is used (includes $\chi(0)$) then these values are strongly modified, indicating mutual dependence of the three fitting parameters. It should be noted that $\Theta_p$ changes drastically if the lower limit of the temperature fitting range is changed.) $\chi$ data for $\bf{H}\|\bf{a}$ show a broad hump at $\sim$ 70 K possibly associated with the CEF effects. The tetragonal crystal field that acts on the Ce$^{3+}$ ion lifts the sixfold degeneracy of the J = 5/2 ground state multiplet resulting in three Kramer’s doublets. The excited crystalline electric field (CEF) levels ($\Delta_1$ and $\Delta_2$) were estimated using a mean field theory model to lie at around 189 and 283 K, respectively\cite{Adriano2015} and well above the magnetic ordering temperature.

Field-dependent magnetization data $M(H)$ for \textbf{H}$\|\textbf{a}$ and \textbf{H}$\|\textbf{c}$ at 1.8 K are shown in Fig. \ref{p1}(b). $M$ versus $H$ for \textbf{H}$\|\textbf{c}$ shows a first-order hysteretic metamagnetic spin-flop transition with a saturated moment of 2.00 $\mu_B$, which is almost equal to that of the Ce$^{3+}$ free ion moment of 2.14$\mu_B$. $M(H)$ for \textbf{H}$\|$\textbf{a} does not show any features up to 140 kOe and is far from reaching the saturated moment of Ce$^{3+}$ indicating that the \textit{a} axis is a hard axis.

Figure \ref{p1}(c) shows the zero field temperature-dependent resistivity $\rho(T)$ data of CeAuBi$_2$. As the temperature decreases, the $\rho(T)$ plot shows a broad shoulder, associated with the thermal depopulation of excited CEF levels, followed by a sharp change in slope and a kink corresponding to the AFM transition at 13.2 $\pm$ 0.2 K.

Heat capacity data $C_p(T)$, Fig. \ref{p1}(d), show a clear second-order AFM transition at 13.2 $\pm$ 0.2 K (an equal entropy construction criterion) and a nuclear Schottky anomaly below 1.3 K. The inset of Fig. \ref{p1}(d) shows $C_p(T)/T$ versus $T^2$. The linear fit of $C_p(T)/T$ versus $T^2$ above 14 K results in $\gamma$ of 572 mJ/mol K$^2$ which signifies a strongly correlated system with enhanced electron mass due to strong electron correlations and a Debye temperature of $\sim$185 K.

\begin{figure}[t]
\centering
\includegraphics[width=0.75\linewidth]{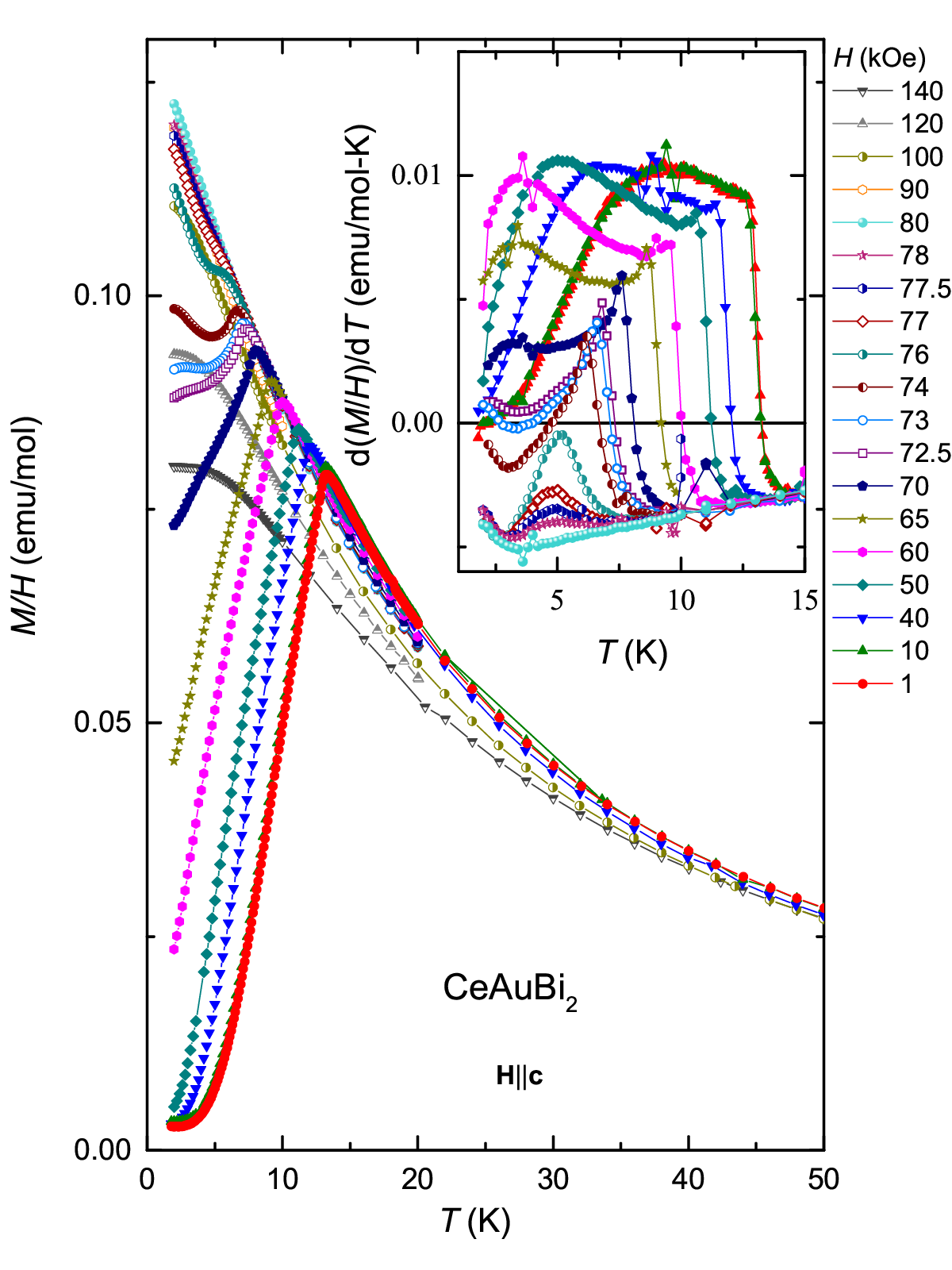}
\caption{\footnotesize Temperature-dependent susceptibility of CeAuBi$_2$ single crystal for \textbf{H}$\|\textbf{c}$. Inset, d($M$/$H$)/d$T$ of CeAuBi$_2$, \textbf{H}$\|\textbf{c}$.}
\label{p2}
\end{figure}

%----------------------------------------------------------------------------------------------------------
\subsection{Measurements with \textbf{H}$\|$\textbf{c}}

Figure \ref{p2} shows the temperature-dependent $M/H$ for \textbf{H}$\|$\textbf{c} measured at constant applied magnetic fields. The kink associated with $T_N$ is suppressed to low temperature with increasing $H$. Starting from $H$ = 72.5 kOe, the peak broadens and additional features appear below $T_N$ upon cooling. This feature may be associated with the spin-flop transition. The inset of Fig. \ref{p2} shows the derivative of the $M/H$ data presented in the main panel of Fig. \ref{p2}. At $H$ = 1 kOe, the $T_N$ transition manifests itself as a sharp increase in d($M/H$)/d$T$ followed by a broad hump when the temperature decreases. Starting from 76 kOe, only a broad maximum that denotes a position of a spin-flop transition is seen. This maximum disappears between 78 and 80 kOe.

\begin{figure}[t]
\centering
\includegraphics[width=1\linewidth]{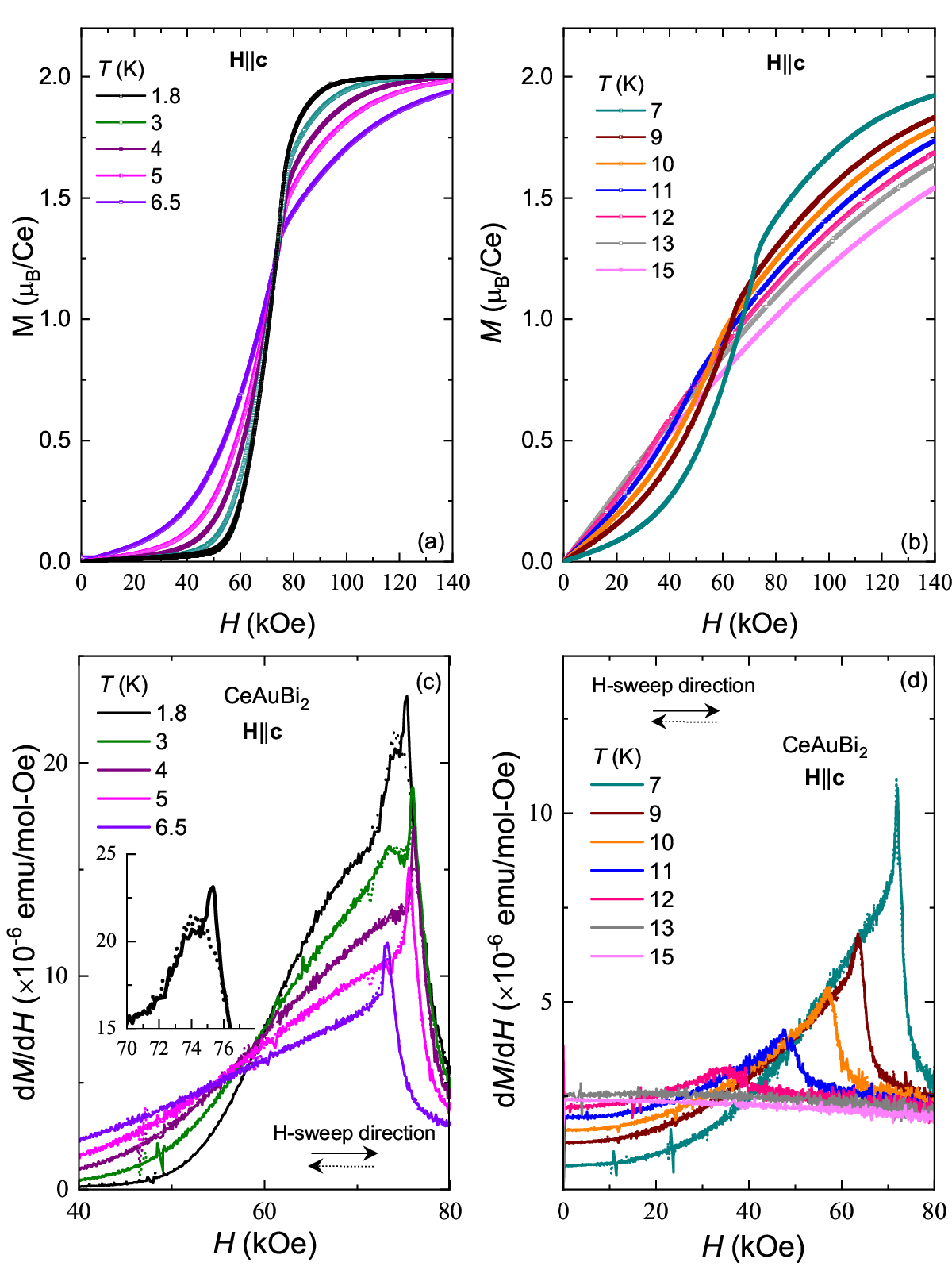}
\caption{\footnotesize (a) and (b) Field-dependent magnetization of CeAuBi$_2$ single crystal for \textbf{H}$\|\textbf{c}$. The curve for $T$ = 6.6 K is shown on both panels for continuity. (c) and (d) the derivative d$M$/d$H$ of the data presented in panels (a) and (b), respectively. Inset of figure (c) shows zoomed in hysteresis region for 1.8 K. }
\label{MH}
\end{figure}

Field-dependent magnetization $M(H)$ at constant temperatures is shown in Figs. \ref{MH}(a) and (b). The data is split into two panels for better data visualization. The $M(H)$ data were collected on the magnetic field sweep up and down for all temperatures below $T$ = 11 K. A small hysteresis is more apparent at $T$ = 1.8 K [better seen in Fig. \ref{p1}(b)] and becomes less distinguishable as the temperature increases.  
In Figures \ref{MH}(c) and (d), we plotted the derivative of the $M(H)$ data presented in Figs. \ref{MH}(a) and (b), respectively. d$M$/d$H$ vs $H$ data show two broad shoulders and a sharp peak between 60-80 kOe at 1.8 K. The hysteresis, apparent in the magnetic field sweep direction, is shown in the inset of Fig. \ref{MH} (c). The data at 3 K, show only one broad and one sharp peak between 70-80 kOe. Starting from $T$ = 4 K, there is only one peak with a small hysteresis at the peak. This peak decreases in amplitude and broadens as the temperature increases, disappearing for $T$ = 13 K, the same temperature as $T_N$.

\begin{figure}[t]
\centering
\includegraphics[width=1\linewidth]{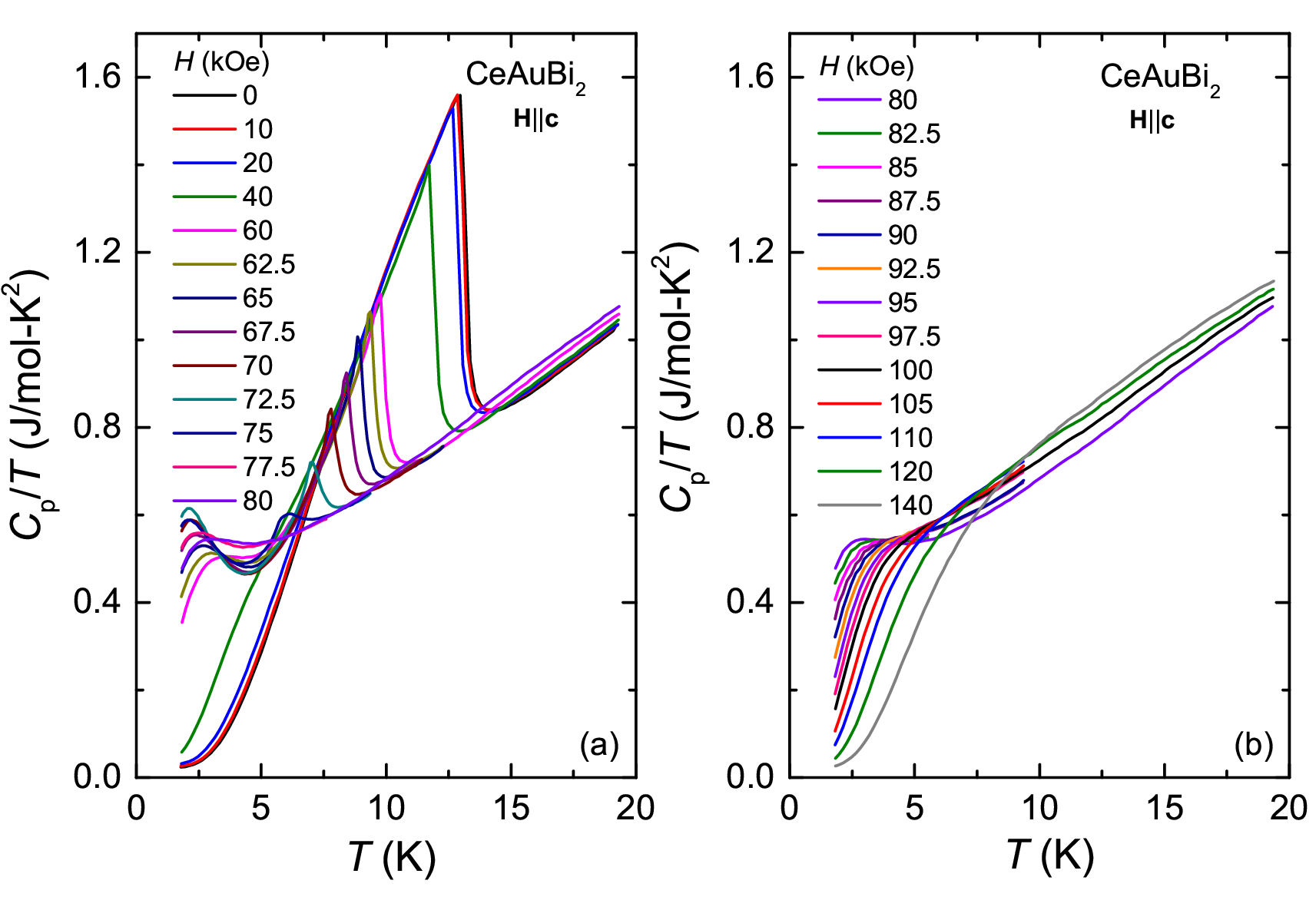}
\caption{\footnotesize (a) and (b) Temperature-dependent heat capacity of CeAuBi$_2$ single crystal $\#$ 1 for \textbf{H}$\|\textbf{c}$. The data curve for $H$ = 80 kOe is shown on both panels for clarity.}
\label{p4}
\end{figure}

The evolution of the $C_p(T)/T$ data at constant magnetic fields is shown in Fig. \ref{p4}. $T_N$ is gradually suppressed with the magnetic field. At $H$ = 65 kOe, the shape of $C_p(T)/T$ at $T_N$ starts to change and evolves into a broad maximum for $H$ = 75 kOe. This broad maximum is no longer observable at $H$ = 77.5 kOe. In addition to $T_N$, starting from $H$ = 60 kOe a broad low-temperature maximum is seen, as the magnetic field is increased, the maximum moves first to lower temperatures, grows in amplitude, reaches highest value at 72.5 kOe, then decreases in amplitude, broadens, and moves to higher temperatures. This behavior of the maximum in the heat capacity, especially for magnetic fields larger than $H$ = 77.5 kOe, is very similar to the evolution of Zeeman splitting of the CEF levels under application of the magnetic field, for example as in CeZn$_{11}$.\cite{Hodovanets2013b} Unfortunately, the position of this maximum is hard to determine precisely beyond $H$ = 85 kOe since we do not have LaAuBi$_2$ to account for the contribution of the phonons.

\begin{figure}[t]
\centering
\includegraphics[width=1\linewidth]{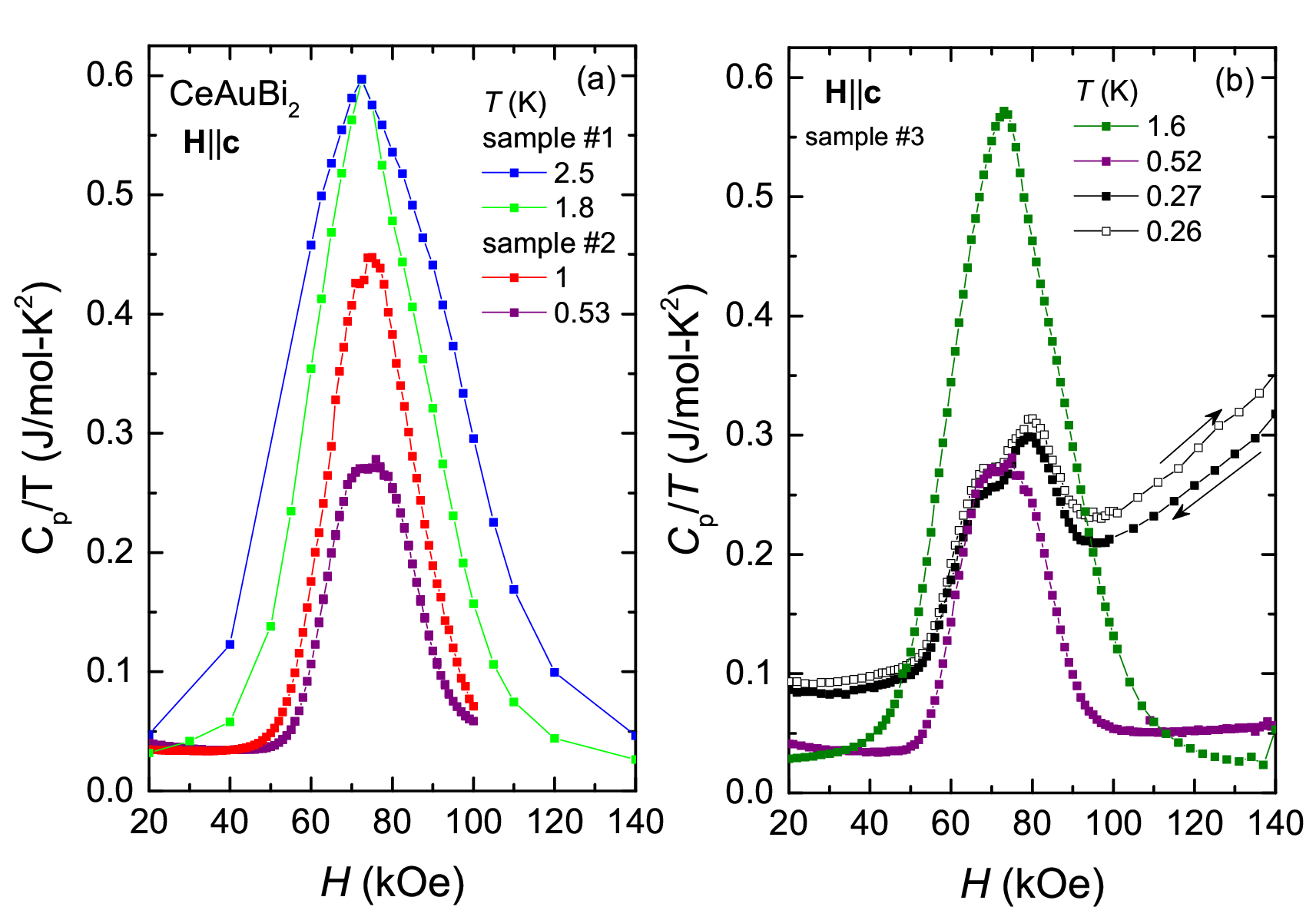}
\caption{\footnotesize Field-dependent $C_p/T$ of CeAuBi$_2$ single crystals (a) samples $\#$1 and  $\#$2, the data for $T$ = 2.5 and 1.8 K are assembled from Fig. \ref{p4}; and (b) sample $\#$3. The arrows denote the sweep direction of the magnetic field for the two lowest temperature data sets. All three samples (single crystals) are from the same batch.}
\label{CvsH}
\end{figure}

\begin{figure*}[t]
\centering
\includegraphics[width=0.7\linewidth]{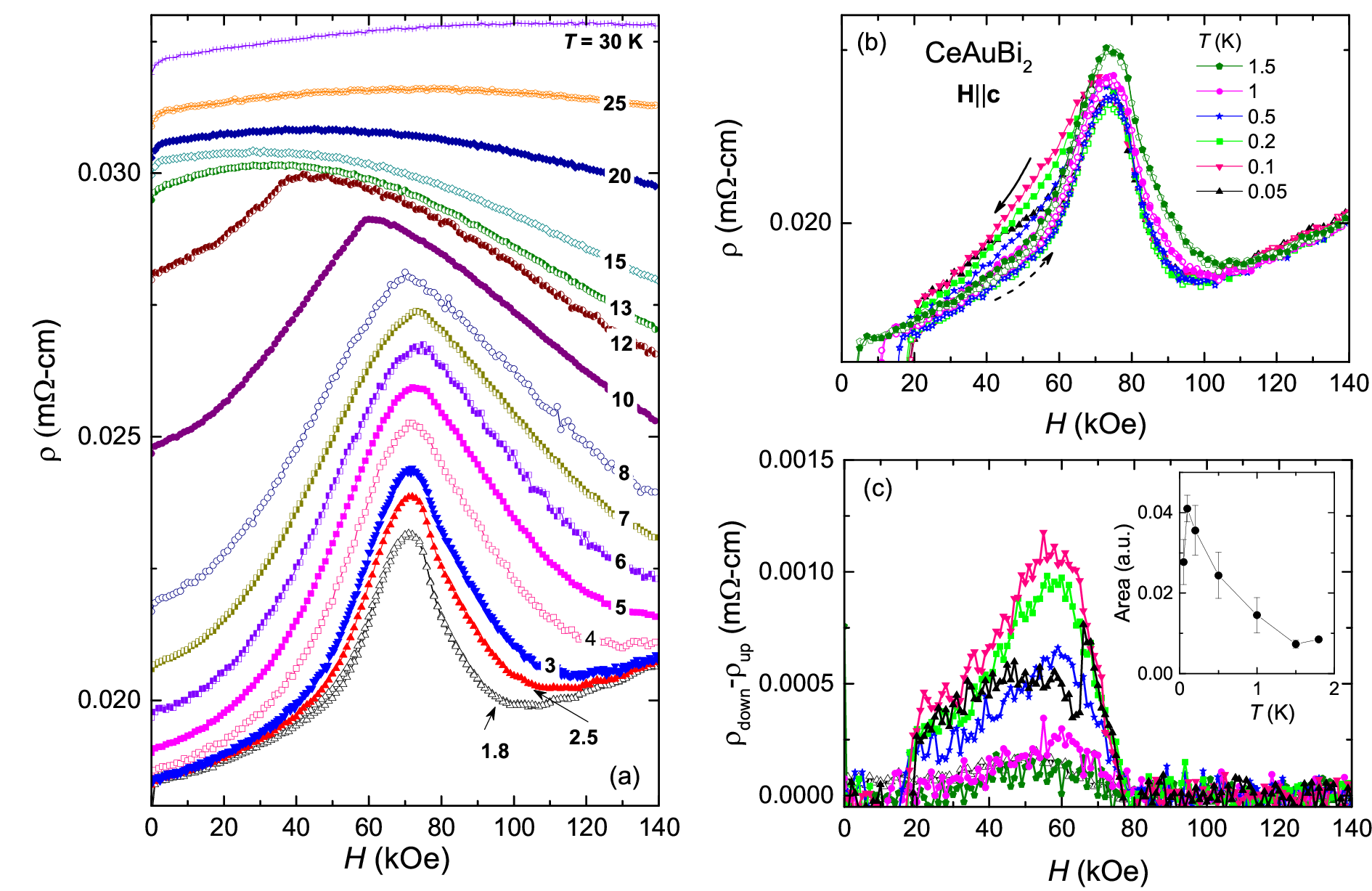}
\caption{\footnotesize Field-dependent resistivity of CeAuBi$_2$ single crystal $\#$ 1 at constant temperatures (a) measured in the PPMS DynaCool and (b) dilution refrigerator (the contacts on the sample were replaced). (c) Difference in $\rho(H)$ data on the field sweep down and up, showing hysteresis. Inset, area of the hysteresis as a function of temperature. \textbf{H}$\|\textbf{c}$}
\label{RoHD}
\end{figure*}

\begin{figure*}[tbh]
\centering
\includegraphics[width=0.7\linewidth]{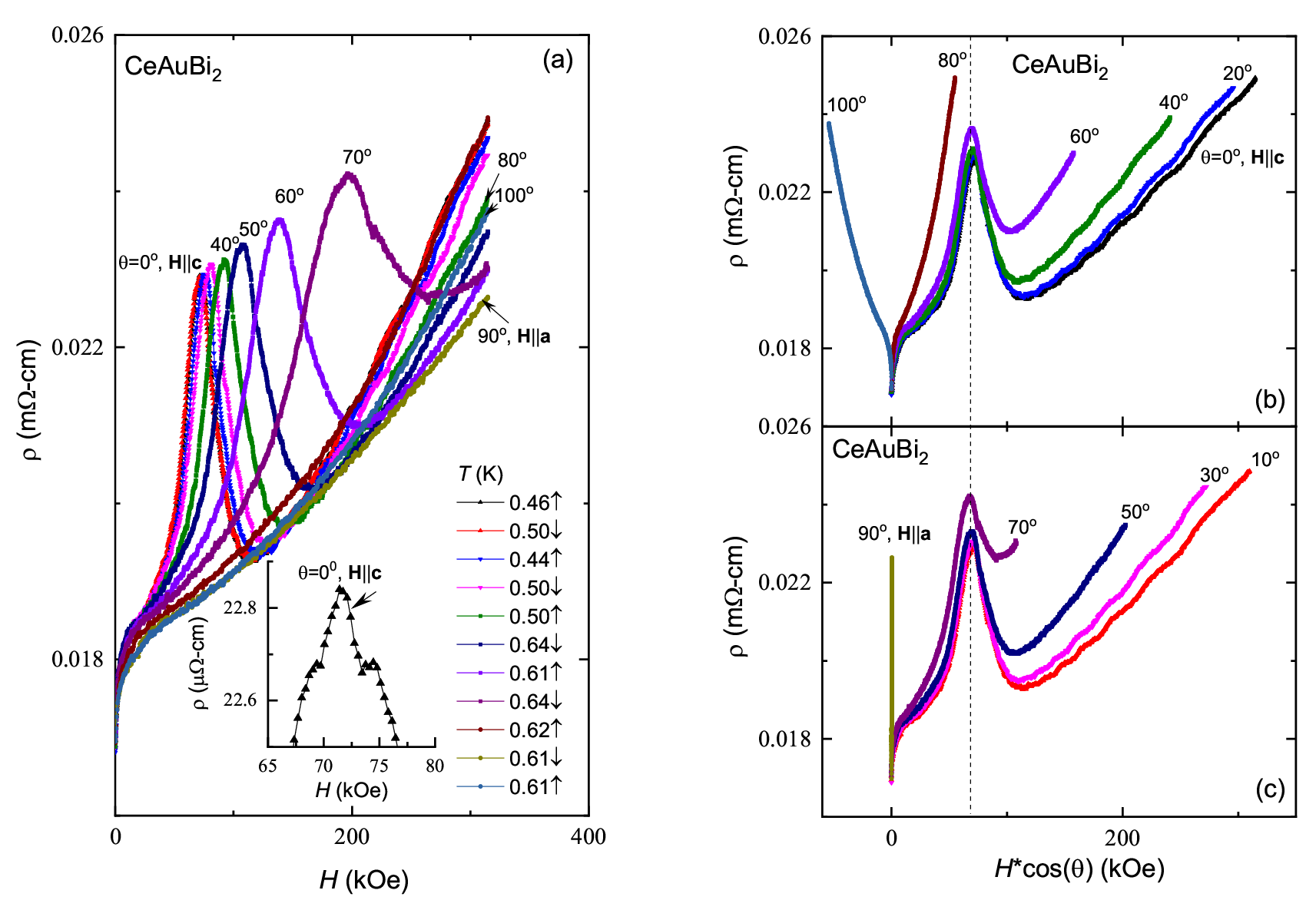}
\caption{\footnotesize (a) $\rho(H)$ of CeAuBi$_2$ for 0$^o$ $\leq\theta\leq$ 110$^o$, where $\theta$ is the angle between the \textit{c}-axis and the applied magnetic field \textit{H}. Inset: $\rho(H)$ for $\theta$ = 0$^o$. $\rho(H)$ as a function of $H\times cos(\theta)$ for (a) magnetic field sweep up and (b) magnetic field sweep down.}
\label{HM}
\end{figure*}

Field-dependent data $C_p(H)/T$, presented in Fig. \ref{CvsH}, show a very large and broad peak centered at $\sim$ 72.5 kOe. As the temperature is lowered, the area of the peak becomes smaller and the peak flattens out. Additional features appear at the peak that evolve into a clear double-peak structure at the lowest temperatures measured. Although the data density is not enough to state with certainty that there are three peaks at $\sim$ 0.5 K in $C_p(H)/T$, the $\rho(H)$ data, shown in the inset of Fig. \ref{HM}(a) below, clearly indicate three peaks at similar magnetic fields. This probably indicates that instead of the single spin-flop transition three metamagnetic transitions are observed at 0.5 K. Despite the clear double transition at $\sim$ 0.27 K, measurements of $M(H)$ are required to see how many metamagnetic transitions are observed as the temperature is lowered below 1.8 K. It should be noted that below $\sim$ 0.5 K the nuclear contribution to the $C_p(H)/T$ data becomes significant. It is most apparent as an increase with magnetic field background for the $T$ = 0.26 and 0.27 K data.

Figure \ref{RoHD} shows the field-dependent resistivity $\rho(H)$ data collected from 30 K to 0.05 K. At 30 K, the magnetoresistance is positive. As the temperature is lowered, the magnetoresistance develops a broad maximum. For 12 K $\leq T\leq$ 7 K, the AFM ordering temperature manifests itself as a kink on a broad maximum. Starting from 6 K and down to the lowest temperature measured, the kink corresponding to AFM ordering is not resolved; instead, a broad maximum centered at $\sim$ 75 kOe is observed in the field-dependent resistivity.

The $\rho(H)$ data for $T\leq$ 1.8 K show hysteresis associated with the first-order transition on the low field side of the maximum, with the resistivity on the field-sweep down being larger than on the field-sweep up. Hysteresis increases in magnitude as the temperature is lowered further to $T$ = 0.05 K. The difference in resistivity between the magnetic field down and up sweeps is shown in Fig. \ref{RoHD}(c). For the lowest temperature measured, no hysteresis in the resistivity is observed above 80 kOe. This field decreases as the temperature increases. The $\rho(H)$ data at 2.5 K (not shown here) did not show measurable hysteresis. Due to the presence of Bi, which manifests itself as a superconducting transition below $\sim$ 20 kOe, no hysteresis in resistivity is observed below this field. The inset of Fig. \ref{RoHD}(c) shows that the area of hysteresis in the resistivity as a function of temperature decreases as the temperature increases to 2 K. The hysteresis in the magnetic field-dependent resistivity above 2 K cannot be resolved within the instrument limitation.

Figure \ref{HM}(a) shows field-dependent resistivity at \textit{T}$\sim$ 0.5 K for 0$^o$ $\leq\theta\leq$ 110$^o$, where $\theta$ is the angle between the \textit{c} axis and the applied magnetic field \textit{H}. When $\theta$ = 0$^o$, the magnetic field is applied along the easy \textit{c} axis and when $\theta$ = 90$^o$, the field is in the \textit{ab}- plane. For $\theta$ = 0$^o$, three peaks/transitions are observed between 65 and 77.5 kOe, inset to Fig. \ref{HM}(a). They are still observable for $\theta$ = 10$^o$. As the angle $\theta$ increases, the transitions move to higher magnetic fields and are eventually not observable in the measured magnetic field range for $\theta\geq$ = 80$^o$. Since the system is Ising-like with the moments aligned along the \textit{c}-axis and thus highly anisotropic, when the sample is rotated with respect to the magnetic field, the perpendicular component of the magnetization along the \textit{c} axis contributes mainly to the $\rho(H)$ data. To test this assumption, we plotted the $\rho(H)$ as a function of $H\times$ cos($\theta$) in Figs. \ref{HM}(b) and (c) for the magnetic field sweep up and down, respectively, due to hysteresis. Indeed, the peak position is unchanged with respect to $\theta$ which confirms our assumption and further supports the Ising-like moment orientation for CeAuBi$_2$.

\begin{figure}[tbh]
\centering
\includegraphics[width=1\linewidth]{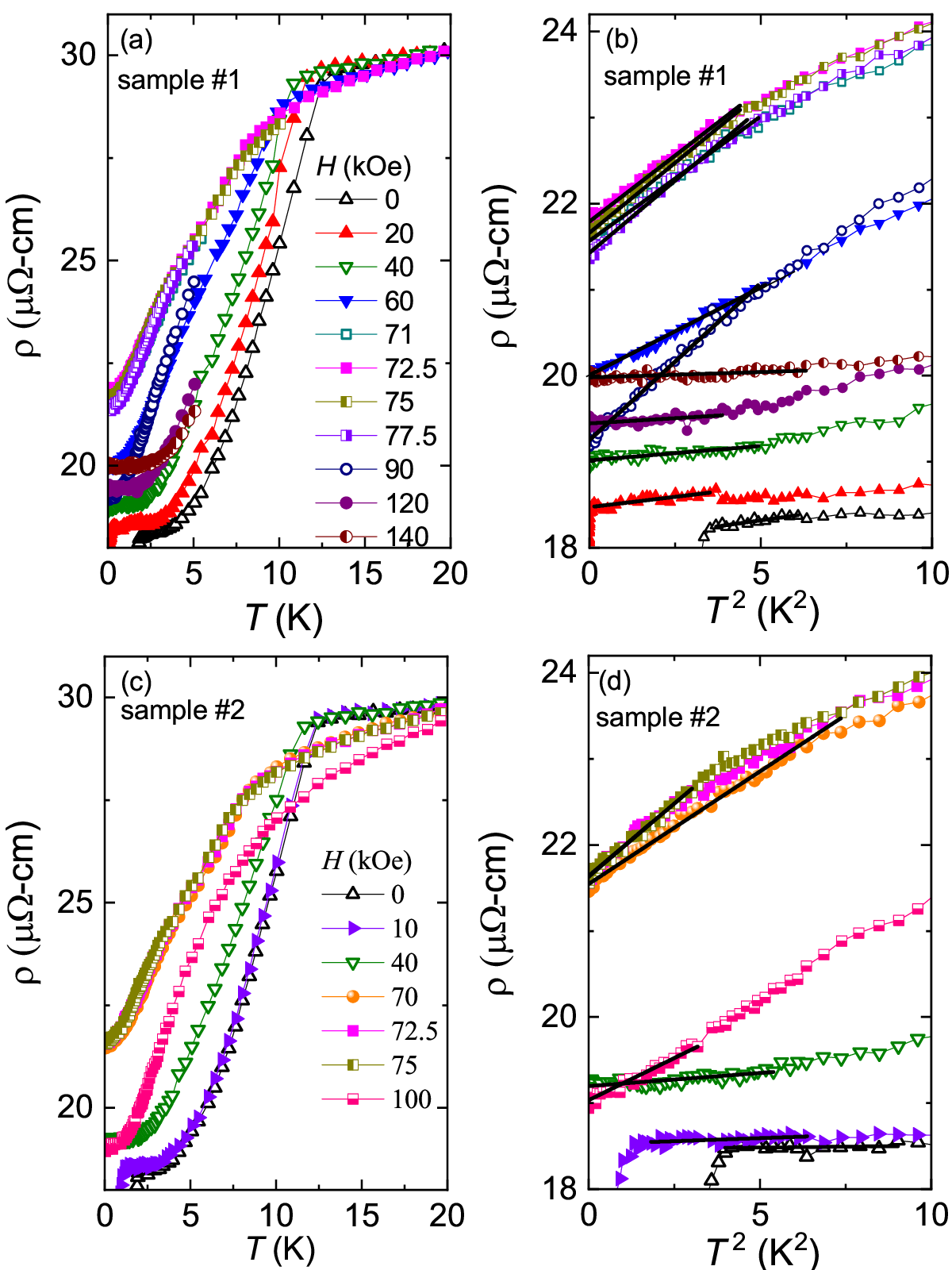}
\caption{\footnotesize Temperature-dependent resistivity of CeAuBi$_2$ two single crystals at constant magnetic field. $\rho$ as a function of (a) $T$ and (b) $T^2$ for sample $\#$ 1 and (c) $T$ and (d) $T^2$ for sample $\#$ 2. \textbf{H}$\|\textbf{c}$. Solid lines represent fitting to $\rho=\rho_0+A\times T^2$. }
\label{RTT}
\end{figure}

To see the functional evolution of the temperature-dependent resistivity below 1.8 K, we extended the $\rho(T)$ data to 0.1 K. Data for two samples are shown in Figs. \ref{RTT}(a) and (c). As the temperature decreases, the sublinear resistivity for $H\sim$ 70 kOe becomes temperature independent. In fact, the resistivity for all fields measured saturates at the lowest temperatures. The data for 0 kOe $\leq H\leq$ 20 kOe show a superconducting transition due to the amorphous Bi. To see the region of Fermi behavior, we plotted $\rho(T)$ data as a function of $T^2$ in Figs. \ref{RTT}(b) and (d). Here, the region of temperatures where the $T^2$ law holds is marked with black lines that follow the fitting lines.

\begin{figure}[ht]
\centering
\includegraphics[width=1\linewidth]{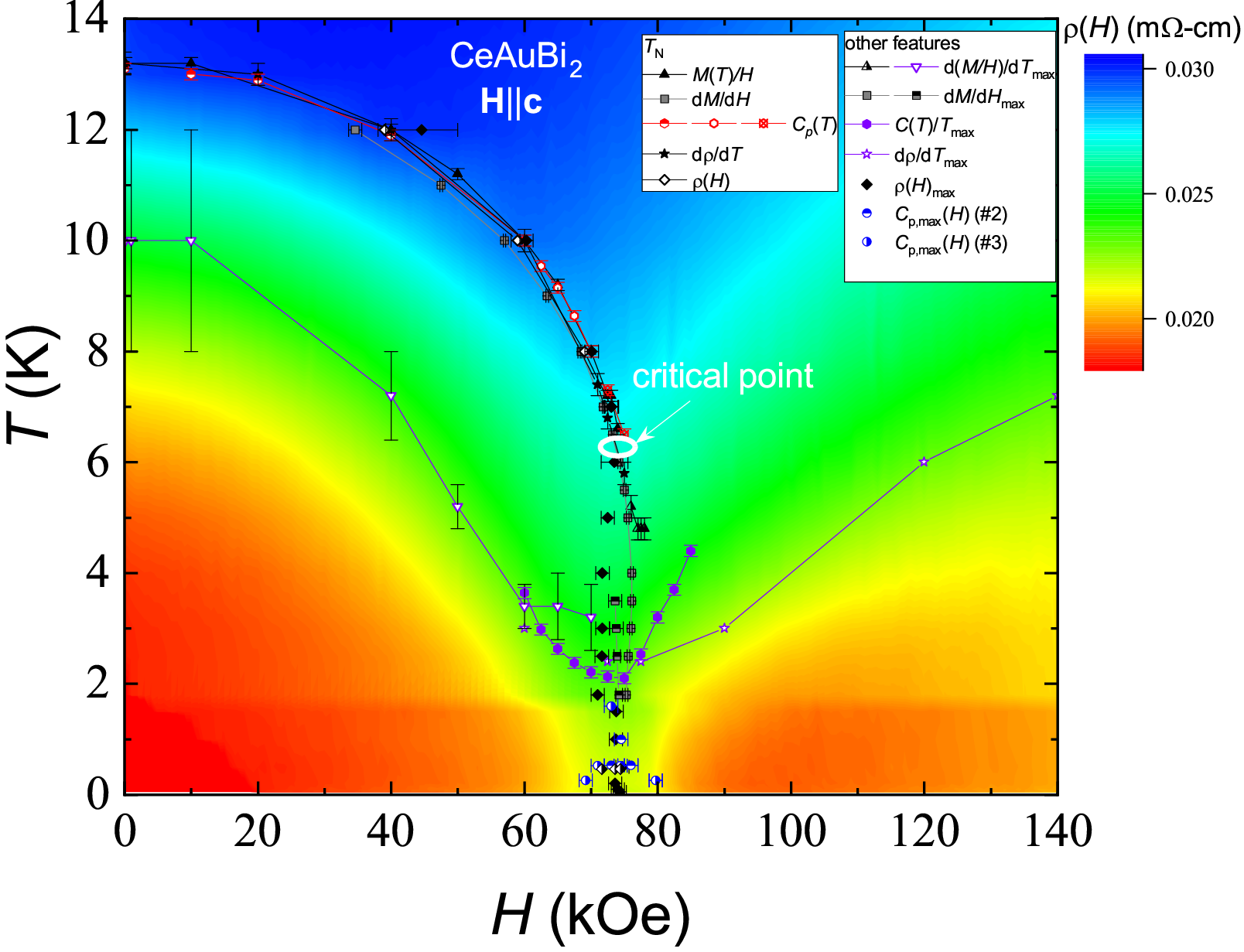}
\caption{\footnotesize $T-H$ phase diagram of CeAuBi$_2$.}
\label{PD}
\end{figure}

\section{Discussion}

The evolution of the AFM transition as a function of the magnetic field, \textbf{H}$\|$\textbf{c} superimposed on the contour plot of $\rho(H)$, is shown in the $T-H$ phase diagram in Fig. \ref{PD}. The slight shift in red color in the contour plot is due to resistivity measurements being taken in two different cryostats. The AFM transition is suppressed by the magnetic field and, below $\sim$ 6.5 K at 75 kOe, evolves into the first-order spin-flop transition. This is highlighted by the tip of the v-shaped light blue region in the contour plot of $\rho(H)$. Upon lowering the temperature below 2 K and increasing the magnetic field, the spin-flop first-order transition line splits into three curves, creating two low-temperature domes. Based on the hysteresis observed in the $\rho(H)$ data shown in Fig. \ref{RoHD}(c), all three field lines are of the first order.

The broad peak below the AFM order in the d($M/H$)/d$T$ on the low-field side of the spin-flop transition and the features in $C(T)/T$ together with d$\rho$/d$T$ data on the high-field side of the spin-flop transition delineate two other lines in the phase diagram. The features in the data that result in these two lines look very similar to those due to the Zeeman splitting of the low-lying CEF levels and would position the lowest excited state at $\sim$ 15 K and the second excited level is at $\sim$ 150-200 K (based on both temperature-dependent magnetization and resistivity measurements). However, the two excited CEF levels obtained based on the fit to the CEF mean field model were estimated to be 189 and 283 K\cite{Adriano2015} or slightly higher values for the stoichiometric CeAuBi$_2$\cite{Piva2020}. As was pointed out in Ref. \cite{Adriano2015}, the fit to the CEF mean field model may not be as precise and unique. Nevertheless, inelastic neutron scattering measurements are required to support this statement. The lower values of the CEF levels are also more in line with those estimated for CeCuBi$_2$ and CeAgBi$_2$. It should be noted that we did not observe any change in the magnetic structure between 4 and 12 K at H=0 in the neutron diffraction measurements.  

\begin{figure*}[tbh]
\centering
\includegraphics[width=0.75\linewidth]{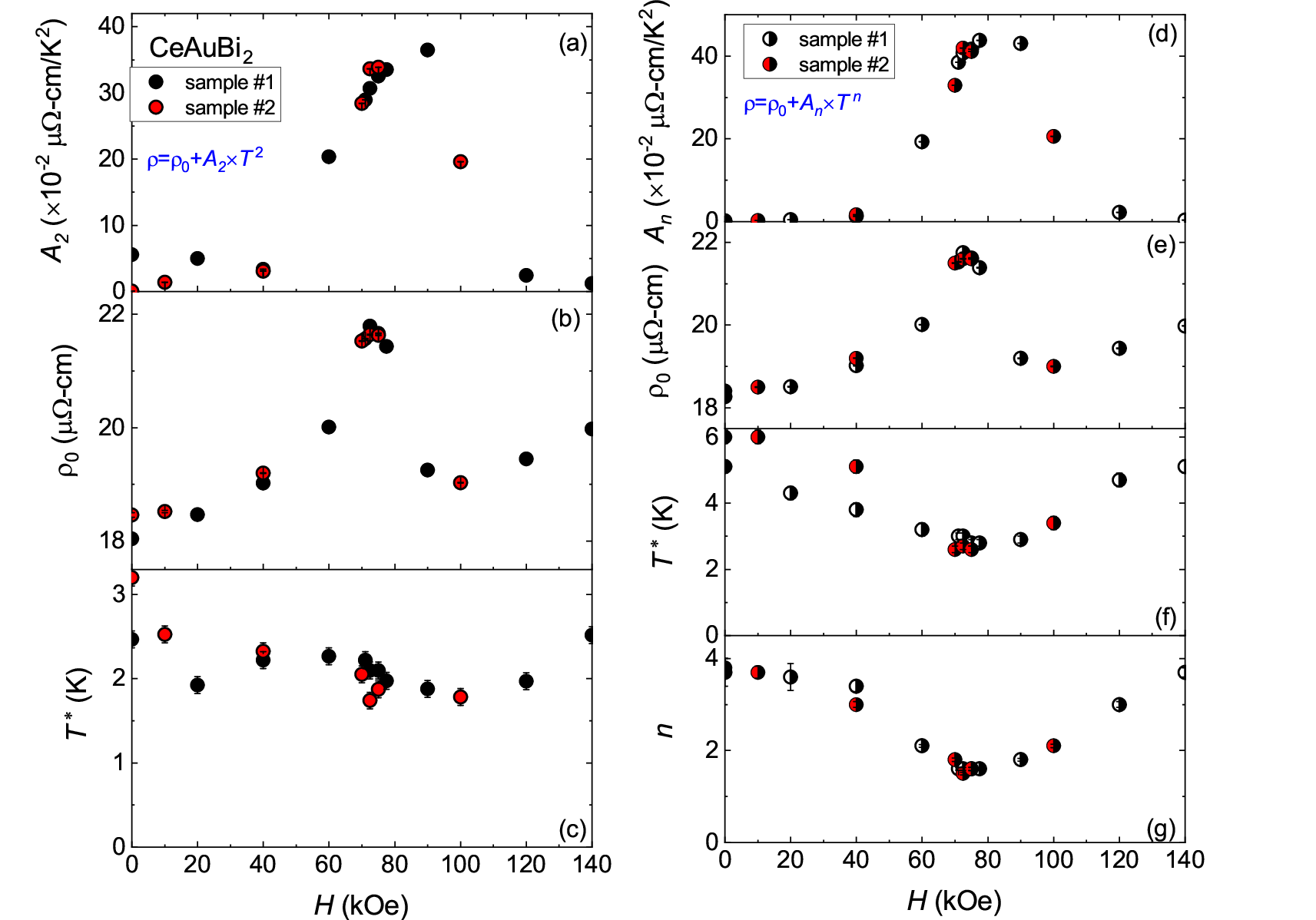}
\caption{\footnotesize Results of the $\rho(T)$ data fits (a), (b), and (c) $\rho=\rho_0+A_2\times T^2$ and (d), (e), (f), and (g) $\rho=\rho_0+A_n\times T^n$. $T^*$ is the upper limit of the temperature range where the fitting was performed.}
\label{AT}
\end{figure*}

To check for signs of critical behavior, we fitted the temperature-dependent resistivity data to the $\rho(T)=\rho(0)+A_n\times T^n$ fit, where the exponent $n$ is 2 (Fig. \ref{RTT}) or a fitting parameter, and the results are presented in Fig. \ref{AT}. For the curves that show the Bi superconducting transition, the fit was done above the superconducting transition. The coefficients $A_2$ and $A_n$ shown in Figs. \ref{AT}(a) and (d) are quite consistent between the two samples. They peak at $H\sim$ 85 kOe. However, they are rather small compared to what is expected for the heavy-fermion systems. $\rho(0)$ shown in Figs.\ref{AT}(b) and (e) peaks at $H\sim$ 75 kOe, and $T^*$, the upper temperature limit of the fit has a small feature at about the same magnetic field. The upper limit $T^*$ of the temperature region where the fit was performed is shown in Figs. \ref{AT}(c) and (f). The $T^*$ stays nearly the same for the quadratic fit. For the fit of the power function, $T^*$ decreases to its smallest value of $\sim$ 2.8 K at 75 kOe, after which it increases again. The exponent $n$, Fig. \ref{AT}(g) shows similar behavior to $T^*$, Fig. \ref{AT}(g), it dips to 1.6 at 75 kOe. $\rho_0$ also shows enhancement at the critical region and so does an electronic coefficient $C_p/T$ vs. $H$ presented in Fig. \ref{CvsH}(b). Although field-induced QCP is avoided, at the lowest temperature $C_p/T$ vs. $H$ is largely enhanced across the region of the two domes and attains almost 300 mJ/mol-K$^2$ at its highest value compared to less than 50 mJ/mol-K$^2$ at 40 and 100 kOe. 
It should be noted that the $T-H$ phase diagram that we present in this work is "much simpler" than that assembled for the stoichiometric CeAuBi$_2$\cite{Piva2020} due to the absence of many metamagnetic transitions observable in the nearly stoichiometric CeAuBi$_2$. 

Natural questions arise as to why the $T-H$ phase diagram for a Au deficient CeAuBi$_2$ is simpler than that of the stoichiometric compound and what role do Au vacancies play. Do they affect magnetic anisotropy and exchange interaction? When we compare $T_N$= 13.2 K for our work with that of nearly stoichiometric CeAuBi$_2$, we can see that it is lower. However, the magnetic anisotropy at $T_N$ and the saturated moment are not affected. In Ref. \cite{Thomas2016} spin model for the Ce local moments built on a magnetic structure of CeCuBi$_2$ (which is the same as that for CeAuBi$_2$) at zero field was introduced to explain the many metamagnetic transitions observed for CeAgBi$_2$. In this model, three competing spin exchanges were assigned between neighboring sites: FM and AFM (or AFM and FM) spin exchanges for two symmetry-distinct neighbors along the \textit{c} axis - $J$ and $J'$, respectively, and a FM spin interaction in the $ab$ plane - $J_\perp$. When all these three parameters and an easy-axis single-ion anisotropy along the $c$-axis were equal, at least six magnetic phases were realized in CeAgBi$_2$. The differences in the $T-H$ phase diagrams among CeCuBi$_2$, CeAgBi$_2$, and CeAuBi$_2$ are expected due to different orbitals (3d or 4s for Cu, 4d or 5s for Ag, and 5d or 6s for Au) participating in a spin superexchange between Ce moments and thus changing the effective superexchange couplings.\cite{Thomas2016} This explains the similarities and slight differences between the $T-H$ phase diagrams of CeAgBi$_2$ and stoichiometric CeAuBi$_2$. However, this model did not take into account nonstoichiometry of the Ag and it's affect on the complexity of the different magnetic phases observed in the $T-H$ phase diagram.

Let us see what role Au vacancies play. To understand the interplay among exchange interactions, single-ion anisotropy, and quenched disorder in CeAuBi$_2$, we adapt the classical spin model put forward by Ref. \cite{Thomas2016} for CeAgBi$_2$. The model Hamiltonian reads
\begin{equation}
H = \sum_{ij} J_{ij} \mathbf{S}_i \cdot \mathbf{S}_j  -  \sum_i \Delta (S_i^z)^2 - \sum_i \mathbf{h} \cdot \mathbf{S}_i ~.
\label{eq:Hamiltonian}
\end{equation} 
Here, the $\mathbf{S}_i$ are classical Heisenberg spins that represent the Ce moments.  $\mathbf{h}$ denotes the external magnetic field, and $\Delta$ is the single-ion anisotropy energy. For $\Delta > 0$, it is of easy-axis type. The $J_{ij}$ represent the exchange interactions between the Ce moments. A minimal model of physics in CeAuBi$_2$ includes interactions $J$ and $J'$ between the two symmetry-distinct pairs of neighbors in the $c$ direction as well as an interaction $J_\perp$ in the $ab$ plane\cite{Thomas2016}.  

The zero-field low-temperature magnetic state of CeAuBi$_2$ (as well as CeAgBi$_2$) consists of ferromagnetic layers arranged in a
$(++--)$ pattern in the $c$ direction. This can be modeled by choosing $J$ to be ferromagnetic, while $J'$ and $J_\perp$ are antiferromagnetic. Quenched disorder is introduced by adding independent random contributions to the interactions $J$, $J'$, and $J_\perp$. They are drawn from Gaussian distributions of zero mean and standard deviations $\delta J$, $\delta J'$, and $\delta J_\perp$, respectively.

To find the (low temperature) magnetic state for a given set of parameters $J_{ij}$, $\Delta$, and $\mathbf{h}$, we minimize the Hamiltonian (\ref{eq:Hamiltonian}) in a magnetic unit cell containing 16 Ce sites (two bilayers). Further details are given in the Experimental Details section.

\begin{figure}[b]
\includegraphics[width=\columnwidth]{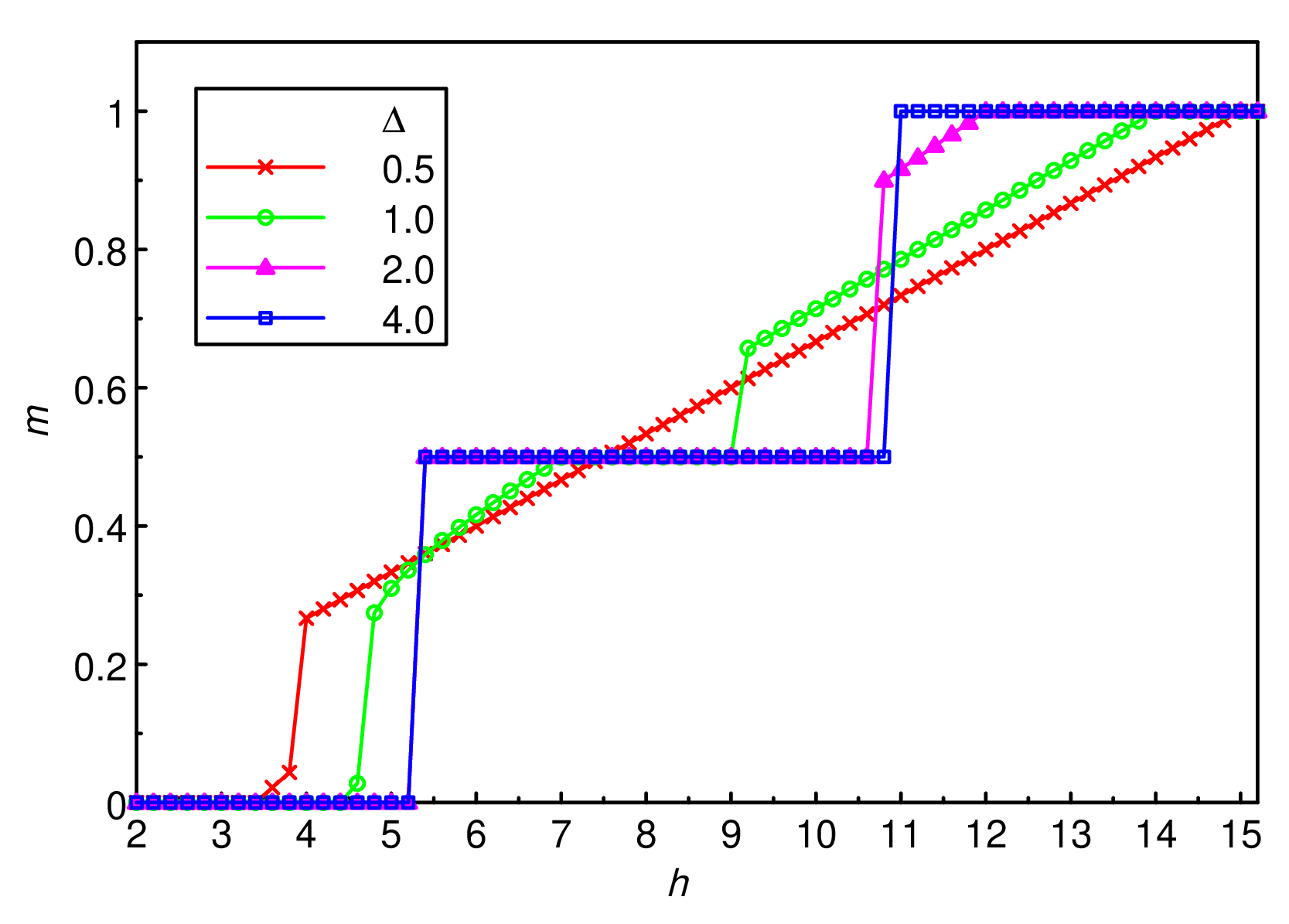}
\caption{Low-temperature magnetization $m$ vs.\ applied field $h$ in the \textit{c} direction for several values of the single-ion anisotropy energy $\Delta.$ The interaction parameters are $J=-1$, $J'=2$, and $J_\perp = 1.2$. There is no quenched disorder, $\delta J= \delta J' = \delta J_\perp=0$.}
\label{fig:anisotropy}
\end{figure}

We start by considering the influence of the single-ion anisotropy in the absence of disorder, $\delta J$=$\delta J'$=$\delta J_\perp$ = 0. CeAuBi$_2$ has a much stronger single-ion anisotropy than CeAgBi$_2$, as signified by the much larger $\chi_c / \chi_a$ ratio. In Fig.\ \ref{fig:anisotropy}, we therefore study the evolution of the magnetization field curve with increasing $\Delta$. 
The values of the interaction energies, $J=-1$, $J'=2$, and $J_\perp = 1.2$, are similar to the ones used for CeAgBi$_2$ in Ref.\cite{Thomas2016}, where they were shown to produce a rich behavior with six different phases. For the weakest single-ion anisotropy, $\Delta=0.5$, we observe a single spin-flop transition at $h \approx 4$, followed by a continuous rotation of the spins until the fully polarized state is reached at $h \approx 15$. For $\Delta=1$ (the value used in Ref. \cite{Thomas2016}), we recover the rich behavior observed for CeAgBi$_2$, including two partial spin-flop transitions and a half-magnetization plateau in which one of the two bilayers is stripe-ordered. For larger $\Delta$, the spin rotations are suppressed, giving way to Ising-like behavior at $\Delta=4$, with a spin-flip transition of the $--$ layers to a stripe state at $h \approx 5.5$ and another partial spin-flip transition from the stripe state to the fully polarized state at $h \approx 11$. 

The single-ion anisotropy energy $\Delta$ for CeAuBi$_2$  can be estimated from a fit of the high-temperature (100 K to 250 K) behavior of $\chi_c / \chi_a$ with the results of the single-spin Hamiltonian $H_s =  -  \sum_i \Delta (S_i^z)^2 - \sum_i \mathbf{h} \cdot \mathbf{S}_i$. This fit gives $\Delta \approx 300 K$, more than an order of magnitude larger than the N\'eel temperature of  CeAuBi$_2$ which sets the scale for the interaction energies. Together with the results of Fig.\ \ref{fig:anisotropy}, this implies that the low-temperature magnetic behavior of CeAuBi$_2$ can be described by an Ising model.

\begin{figure}[tbh]
\includegraphics[width=\columnwidth]{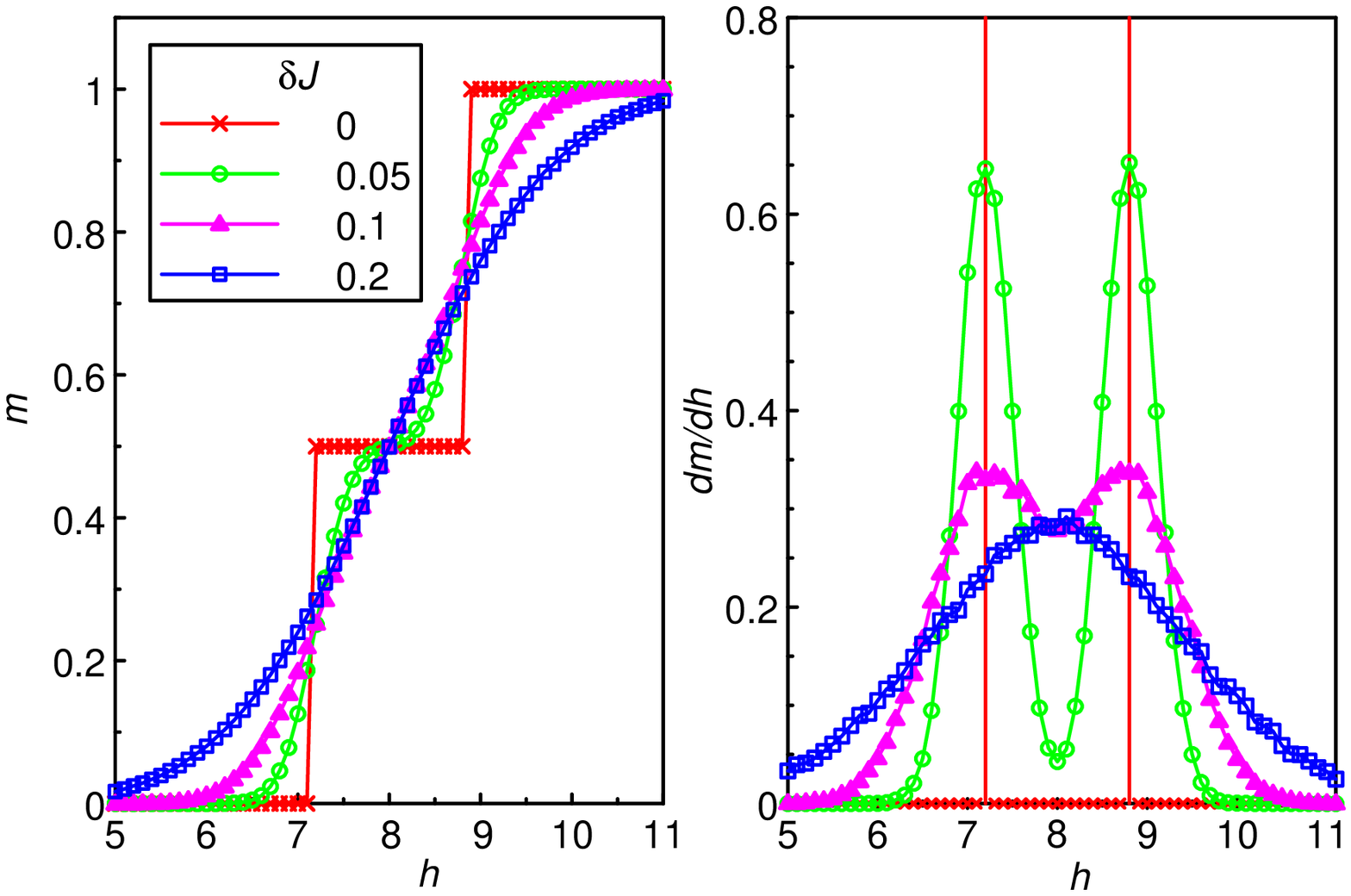}
\caption{Left: Low-temperature magnetization $m$ vs.\ applied field $h$ in the c direction for several values of the disorder strength. 
The interaction energies are drawn from Gaussian distributions with averages $J=-2$, $J'=2$, and $J_\perp = 1.2$ and standard deviation
$\delta J= \delta J' = \delta J_\perp$. 
The results are for the Ising  limit, $\Delta = \infty$. They represent averages over $10^4$ to $10^5$ independent realizations of the 16-spin
magnetic unit cell. Right: $dm/dh$ vs.\ $h$ for the same data sets.}
\label{fig:disorder}
\end{figure}

We now turn to the effects of quenched disorder. Figure \ref{fig:disorder} shows the evolution of the magnetization field curves (in the Ising limit $\Delta = \infty$) with increasing disorder strength $\delta J$=$\delta J'$=$\delta J_\perp$.

The figure shows that the half-magnetization plateau becomes a shoulder-like feature already at $\delta J =0.05$. For $\delta J =0.2$ (still much smaller than the average interactions), this feature is completely suppressed, and $dm/dh$ shows a single broad peak instead of two sharp peaks corresponding to the spin-flip transitions observed in the clean case. We thus conclude that weak quenched disorder is sufficient to smear the complex structure of the $m-h$ curves realized in clean samples. With respect to stabilization of the many metamagnetic transitions and complexity of the in-field magnetic structures of the CeAuBi$_2$ compound, the Au deficiency can be used as a tuning parameter, in addition to pressure and chemical substitution, to tune the system to the QCP.

\section{Conclusion}
We presented detailed thermodynamic and transport measurements in the magnetic field applied along the easy tetragonal \textit{c}-axis collected on single crystals of non-stoichiometric CeAu$_{1-x}$Bi$_2$, \textit{x}=0.18. The compound studied orders antiferromagnetically below $\sim$ 13 K with magnetic moments aligned along the easy tetragonal \textit{c}-axis, Ising system. CeAuBi$_2$ has the highest $\chi_c/\chi_a$ ratio in the Ce$TX_2$ (where $T$ = Cu, Ag, and Au, and $X$ = Sb and Bi) family. The assembled $T-H$ phase diagram shows fewer stabilized magnetic structures compared to those of other members in the Ce$TX_2$ family, particularly stoichiometric CeAuBi$_2$. Our proposed theoretical model shows that weak quenched disorder (Au deficiencies in this case) is sufficient to smear the complex structure of the $M$($H$) curves, implying that Au deficiencies can be used as a tuning parameter. 
%%%%%%%%%%%%%%%%%%%%%%%%%%%%%%%%%%%%%%%%%%%%%%%%%%%%%%%%%%%%%%%%%%%%%%%%%%%%%%%%%%%%%%%%

\section{Acknowledgment} 
The authors thank C. M. Brown and I. Liu for helping with the neutron data collection; C.M Brown, Jacob Mizeur, E. E. Rodriguez, and S. M. Disseler for helping with the neutron diffraction refinement; J. Higgins for helping with the data measurements in the DynaCool, R. Fernandes for fruitful discussions.
This work was supported by the Gordon and Betty Moore Foundation’s EPiQS Initiative through Grant No. GBMF9071, the U.S. National Science Foundation (NSF) Grant No. DMR2303090, and the Maryland Quantum Materials Center.
S.R.S. acknowledges support from the National Institute of Standards and Technology Cooperative Agreement 70NANB17H301. 
A portion of this work was performed at the National High Magnetic Field Laboratory, which is supported by the National Science Foundation Cooperative Agreement No. DMR-1157490 and the State of Florida.

\end{document}